\documentclass[]{aastex701} %linenumbers,[trackchanges removed or ArXiv

\usepackage{gensymb}
\usepackage{amssymb}
\begin{document}

\title{Characterization of the Host Binary of the Directly Imaged Exoplanet HD~143811~AB~b}

\author[orcid=0000-0003-2461-6881]{Anne E. Peck}
\affiliation{Department of Astronomy, New Mexico State University, P.O. Box 30001, MSC 4500, Las Cruces, NM 88003, USA}
\email[show]{annepeck@nmsu.edu}  

\author[orcid=0009-0008-9687-1877]{William Roberson}
\affiliation{Department of Astronomy, New Mexico State University, P.O. Box 30001, MSC 4500, Las Cruces, NM 88003, USA}
\email[]{wcr@nmsu.edu}  

\author[orcid=0000-0001-6975-9056]{Eric L. Nielsen}
\affiliation{Department of Astronomy, New Mexico State University, P.O. Box 30001, MSC 4500, Las Cruces, NM 88003, USA}
\email[]{nielsen@nmsu.edu}  

\author[orcid=0000-0002-4918-0247]{Robert J. De Rosa}
\affiliation{European Southern Observatory, Alonso de C\'{o}rdova 3107, Vitacura, Casilla 19001, Santiago, Chile}
\email[]{rderosa@eso.org}  

\author[orcid=0000-0002-4682-7797]{Nathalie Jones}
\affiliation{Department of Physics and Astronomy, Northwestern University, 2145 Sheridan Road, Evanston, IL 60208-3112, USA}
\affiliation{Center for Interdisciplinary Exploration and Research in Astrophysics, 1800 Sherman Ave,
Northwestern University, Evanston, IL 60201, USA}
\email[]{nathaliejones2028@u.northwestern.edu}  

\author[orcid=0000-0003-0774-6502]{Jason Wang}
\affiliation{Department of Physics and Astronomy, Northwestern University, 2145 Sheridan Road, Evanston, IL 60208-3112, USA}
\affiliation{Center for Interdisciplinary Exploration and Research in Astrophysics, 1800 Sherman Ave,
Northwestern University, Evanston, IL 60201, USA}
\email[]{jason.wang@northwestern.edu }  

\author[0000-0003-1212-7538]{Bruce A. Macintosh} \affiliation{Department of Astronomy and Astrophysics, UC Santa Cruz, Santa Cruz CA 95064} \email[]{bamacint@ucsc.edu}

\author[orcid=0000-0002-8984-4319]{Briley L. Lewis}
\affiliation{Department of Physics, University of California, Santa Barbara, CA 93106, USA}
\email[]{brileylewis@ucsb.edu} 

\author[0000-0002-5092-6464]{Gaspard Duch\^ene}
\affiliation{Department of Astronomy, University of California, Berkeley, CA 94720, USA}
\affiliation{Univ. Grenoble Alpes/CNRS, IPAG, F-38000 Grenoble, France}
\email[]{gduchene@berkeley.edu} 

\author[0000-0003-3050-8203]{Stanimir Metchev}
\affiliation{Department of Physics \& Astronomy, Institute for Earth and Space Exploration, The University of Western Ontario, London, ON N6A 3K7, Canada.}
\email[]{smetchev@uwo.ca} 

\author[0009-0006-4626-832X]{Asif Abbas}
\affiliation{Department of Astronomy, New Mexico State University, P.O. Box 30001, MSC 4500, Las Cruces, NM 88003, USA}
\email[]{asif@nmsu.edu}

\author[orcid=0000-0002-6618-1137]{Jerry W. Xuan}
\affiliation{Department of Astronomy, California Institute of Technology, Pasadena, CA 91125, USA}
\email[]{jxuan@astro.caltech.edu} 

\author[0000-0002-1838-4757]{Aniket Sanghi}
\altaffiliation{NSF Graduate Research Fellow}
\affiliation{Cahill Center for Astronomy and Astrophysics, California Institute of Technology, 1200 E. California Boulevard, MC 249-17, Pasadena, CA 91125, USA}
\email[]{asanghi@caltech.edu}

\author[0000-0001-9004-803X]{Jennifer Patience}
\affiliation{School of Earth and Space Exploration, Arizona State University, Tempe, AZ 85287, USA}
\email[]{jpatienc@asu.edu}

%\author[0000-0001-5172-7902]{S. Mark Ammons} \affiliation{Lawrence Livermore National Laboratory, 7000 East Avenue, Livermore, CA 94550} \email[]{ammons1@llnl.gov}

%\author[0000-0002-5407-2806]{Vanessa P. Bailey} \affiliation{Jet Propulsion Laboratory, California Institute of Technology, 4800 Oak Grove Drive, Pasadena, CA 91109, USA} \email[]{vanessa.bailey@jpl.nasa.gov}

\author[0000-0002-7129-3002]{Travis S. Barman} \affiliation{Lunar and Planetary Lab, University of Arizona, Tucson, AZ 85721, USA} \email[]{barman@lpl.arizona.edu}

\author[0000-0003-4641-2003]{Joanna Bulger} \affiliation{Institute for Astronomy, University of Hawai’i, 2680 Woodlawn Drive, Honolulu, HI 96822, USA} \email[]{jbulger@hawaii.edu}

%\author[0000-0002-6246-2310]{Eugene Chiang} \affiliation{Department of Astronomy, 501 Campbell Hall, University of California Berkeley, Berkeley, CA 94720-3411, USA} \email[]{echiang@astro.berkeley.edu}

\author[0000-0001-6305-7272]{Jeffrey K. Chilcote} \affiliation{Department of Physics and Astronomy, University of Notre Dame, 225 Nieuwland Science Hall, Notre Dame, IN, 46556, USA} \email[]{jchilcote@nd.edu}

\author[0000-0002-0792-3719]{Thomas M. Esposito} \affiliation{Department of Astronomy, 501 Campbell Hall, University of California Berkeley, Berkeley, CA 94720-3411, USA} \affiliation{SETI Institute, Carl Sagan Center, 339 Bernardo Ave Ste 200, Mountain View, CA 94043, USA} \email[]{tesposito@berkeley.edu}

\author[0000-0002-0176-8973]{Michael P. Fitzgerald} \affiliation{Department of Physics \& Astronomy, University of California, Los Angeles, CA 90095, USA} \email[]{mpfitz@ucla.edu}

\author[0000-0002-7821-0695]{Katherine B. Follette} \affiliation{Physics and Astronomy Department, Amherst College, 25 East Drive, Amherst, MA 01002, USA} \email[]{kfollette@amherst.edu}

\author[0009-0000-8603-169X]{Hannah Gallamore} \affiliation{Department of Astronomy, New Mexico State University, P.O. Box 30001, MSC 4500, Las Cruces, NM 88003, USA}
\email[]{hjgallamore@gmail.com}

\author[0000-0002-4144-5116]{Stephen Goodsell} \affiliation{Department of Physics, Durham University, Stockton Road, Durham DH1, UK} \affiliation{Gemini Observatory, Casilla 603, La Serena, Chile} \email[]{stephen.goodsell@noirlab.edu}

\author[]{James R. Graham} \affiliation{Department of Astronomy, 501 Campbell Hall, University of California Berkeley, Berkeley, CA 94720-3411, USA} \email[]{jrg@berkeley.edu}

\author[0000-0002-7162-8036]{Alexandra Z. Greenbaum} \affiliation{IPAC, Mail Code 100-22, Caltech, 1200 E. California Blvd., Pasadena, CA 91125, USA} \email[]{azg@ipac.caltech.edu}

\author[0000-0003-3726-5494]{Pascale Hibon} \affiliation{European Southern Observatory, Alonso de C\'{o}rdova 3107, Vitacura, Casilla 19001, Santiago, Chile} \email[]{phibon@eso.org}

\author[0000-0003-3715-8138]{Patrick Ingraham} \affiliation{Vera C. Rubin Observatory, 950 N Cherry Ave, Tucson AZ, 85719, USA} \email[]{pingraham@lsst.org}

\author[0000-0002-6221-5360]{Paul Kalas} \affiliation{Department of Astronomy, 501 Campbell Hall, University of California Berkeley, Berkeley, CA 94720-3411, USA} \affiliation{SETI Institute, Carl Sagan Center, Mountain View, CA 94043, USA} \affiliation{SETI Institute, Carl Sagan Center, 339 Bernardo Ave Ste 200, Mountain View, CA 94043, USA} \affiliation{Institute of Astrophysics, FORTH, GR-71110 Heraklion, Greece} \email[]{kalas@berkeley.edu}

\author[0000-0002-9936-6285]{Quinn M. Konopacky} \affiliation{Department of Astronomy \& Astrophysics, University of California San Diego, La Jolla, CA, USA} \email[]{qkonopacky@ucsd.edu}

\author[0000-0001-7016-7277]{Franck Marchis} \affiliation{SETI Institute, Carl Sagan Center, 339 Bernardo Ave Ste 200, Mountain View, CA 94043, USA} \email[]{fmarchis@seti.org}

\author[]{J\'{e}r\^{o}me Maire} \affiliation{Department of Astronomy \& Astrophysics, University of California San Diego, La Jolla, CA, USA} \email[]{jmaire@ucsd.edu}

%\author[0000-0002-5251-2943]{Mark S. Marely} \affiliation{Department of Planetary Sciences and Lunar and Planetary Laboratory, University of Arizona, Tucson, AZ 85721, USA} \email[]{marksmarley@arizona.edu}

\author[0000-0002-4164-4182]{Christian Marois} \affiliation{National Research Council of Canada Herzberg, 5071 West Saanich Rd, Victoria, BC, V9E 2E7, Canada} \affiliation{Department of Physics \& Astronomy, University of Victoria, 3800 Finnerty Rd., Victoria, BC V8P 5C2, Canada} \email[]{christian.marois@nrc-cnrc.gc.ca}

\author[0000-0003-3017-9577]{Brenda Matthews} \affiliation{Herzberg Astronomy and Astrophysics, National Research Council of Canada, 5071 West Saanich Rd., Victoria, BC V9E 2E7, Canada} 
\affiliation{Department of Physics \& Astronomy, University of Victoria, 3800 Finnerty Rd., Victoria, BC V8P 5C2, Canada} \email{bcmatthews.herzberg@gmail.com}

\author[]{Dimitri Mawet} \affiliation{Department of Astronomy, California Institute of Technology, Pasadena, CA 91125, USA} \affiliation{Jet Propulsion Laboratory, California Institute of Technology, 4800 Oak Grove Drive, Pasadena, CA 91109, USA} \email[]{dmawet@astro.caltech.edu}

\author[0000-0001-6205-9233]{Maxwell A. Millar-Blanchaer}
\affiliation{Department of Physics, University of California, Santa Barbara, CA 93106, USA}
\email[]{maxmb@ucsb.edu} 

\author[0000-0001-7130-7681]{Rebecca Oppenheimer} \affiliation{American Museum of Natural History, Department of Astrophysics, Central Park West at 79th Street, New York, NY 10024, USA} \email[] {bro@amnh.org}

\author[]{David W. Palmer} \affiliation{Lawrence Livermore National Laboratory, 7000 East Avenue, Livermore, CA, 94550, USA} \email[]{dwpalmer10@gmail.com}

\author[0000-0002-3191-8151]{Marshall D. Perrin}
\affiliation{Space Telescope Science Institute, 3700 San Martin Drive, Baltimore, MD 21218, USA}
\email[]{mperrin@stsci.edu} 

\author[]{Lisa Poyneer}

\affiliation{Lawrence Livermore National Laboratory, 7000 East Avenue, Livermore, CA, 94550, USA}
\email[]{poyneer1@llnl.gov} 

\author[]{Laurent Pueyo}

\affiliation{Space Telescope Science Institute, 3700 San Martin Drive, Baltimore, MD 21218, USA}
\email[]{pueyo@stsci.edu} 

\author[0000-0002-9246-5467]{Abhijith Rajan} \affiliation{Space Telescope Science Institute, 3700 San Martin Drive, Baltimore, MD 21218, USA} \email[]{arajan@stsci.edu}

\author[0000-0003-0029-0258]{Julien Rameau}
\affiliation{Trottier Institute for Research on Exoplanets, Université de Montréal, Département de Physique, C.P. 6128 Succ. Centre-ville, Montréal, QC H3C 3J7, Canada}
\affiliation{University of Grenoble Alpes, CNRS, IPAG, F-38000 Grenoble, France}
\email[]{jul.rameau@gmail.com} 

\author[0000-0002-9667-2244]{Fredrik T. Rantakyr\"o} \affiliation{Gemini Observatory, Casilla 603, La Serena, Chile} \email[]{frantaky@gemini.edu}

\author[0000-0003-1698-9696]{Bin Ren} \affiliation{Observatoire de la Côte d’Azur, 96 Bd de l’Observatoire, 06304 Nice, France} \email[]{bin.ren@oca.eu}

\author[0000-0003-2233-4821]{Jean-Baptiste Ruffio} \affiliation{Department of Astronomy \& Astrophysics, University of California San Diego, La Jolla, CA, USA} \email[]{jruffio@ucsd.edu}

\author[0000-0002-8711-7206]{Dmitry Savransky} \affiliation{Sibley School of Mechanical and Aerospace Engineering, Cornell University, Ithaca, NY 14853, USA} \affiliation{Propulsion Laboratory, California Institute of Technology, 4800 Oak Grove Dr., Pasadena, CA 91109, USA} \email[]{ds264@cornell.edu}

\author[0000-0002-6294-5937]{Adam C. Schneider} \affiliation{United States Naval Observatory, Flagstaff Station, 10391 West Naval Observatory Road, Flagstaff, AZ 86005, USA;  Department of Physics and Astronomy, George Mason University, MS3F3, 4400 University Drive, Fairfax, VA 22030, USA} \email[]{aschneid10@gmail.com}

\author[0000-0003-1251-4124]{Anand Sivaramakrishnan} \affiliation{Space Telescope Science Institute, 3700 San Martin Drive, Baltimore, MD 21218, USA} \email[]{anand@stsci.edu}

\author[0000-0002-9156-9651]{Adam J. R. W. Smith}
\affiliation{Department of Astronomy, New Mexico State University, P.O. Box 30001, MSC 4500, Las Cruces, NM 88003, USA}
\email[]{adjsmith@nmsu.edu}

\author[0000-0002-5815-7372]{Inseok Song} \affiliation{Department of Physics and Astronomy, University of Georgia, Athens, GA 30602, USA} \email[]{song@uga.edu}

\author[0000-0003-2753-2819]{Remi Soummer}
\affiliation{Space Telescope Science Institute, 3700 San Martin Drive, Baltimore, MD 21218, USA}
\email[]{soummer@stsci.edu} 

\author[0000-0002-9121-3436]{Sandrine Thomas}
\affiliation{Vera C. Rubin Observatory, 950 N Cherry Ave, Tucson AZ, 85719, USA}
\email[]{sthomas@lsst.org} 

\author[0000-0002-4479-8291]{Kimberly Ward-Duong}
\affiliation{Department of Astronomy, Smith College, Northampton, MA, 01063, USA}
\email[]{kwardduong@smith.edu} 

\author[0000-0002-9977-8255]{Schuyler G. Wolff}
\affiliation{Steward Observatory, University of Arizona, Tucson, AZ 85721, USA}
\email[]{sgwolff@arizona.edu}

\begin{abstract}

HD~143811~AB is the host star to the directly imaged planet HD~143811~AB~b, which was recently discovered using data from the Gemini Planet Imager and Keck NIRC2. A member of the Sco-Cen star-forming region with an age of $13 \pm 4$ Myr, HD~143811~AB is somewhat rare among hosts of directly imaged planets as it is a close stellar binary, with an $\sim$18 day period. Accurate values for the orbital and stellar parameters of this binary are needed to understand the formation and evolutionary history of the planet in orbit. We utilize archival high-resolution spectroscopy from FEROS on the MPG/ESO 2.2-meter telescope to fit the orbit of the binary, and combine with unresolved photometry to derive the basic stellar properties of the system. From the orbit, we derive precise values of orbital period of $18.59090 \pm 0.00007$ days, and mass ratio of $0.886 \pm 0.003$. When combined with stellar evolutionary models, we find masses of both components of $M_A = 1.30^{+0.03}_{-0.05}$ M$_\odot$ and $M_B = 1.15^{+0.03}_{-0.04}$ M$_\odot$. While the current data are consistent with the planet and stellar orbits being coplanar, the 3D orientations of both systems are currently poorly constrained, with additional observations required to more rigorously test for coplanarity.

\end{abstract}


\section{Introduction} 

A vital part of studying exoplanets is understanding their stellar hosts. The key parameter of age, especially important for self-luminous planets, is more readily measured from the host star than the planet, especially if the star is part of a moving group or association (e.g. \citealt{lagrange:2010}, \citealt{rameau:2013}, \citealt{gagne:2018}). The mass of the star is also of great interest, both to constrain orbital motion (e.g. \citealt{derosa:2015}) and to determine how exoplanet occurrence rate depends on stellar mass (e.g. \citealt{johnson:2010, nielsen:2019}).

Exoplanets in binary (and other multiple) systems represent a particularly interesting dataset. Typically, single stars are targeted by exoplanet surveys given observational constraints; however, some binary configurations still allow for planet detection. In the case of direct imaging, exoplanets have been identified orbiting a single star in a multiple system (e.g 51 Eri b, \citealt{macintosh:2015}) as well as circumbinary planets orbiting far from an inner binary (e.g. HD 106906 b, \citealt{bailey:2014}). In addition to providing constraints on planet formation (e.g. \citealt{kraus:2016}), these planetary systems are also more likely to experience complex 3-body dynamics, such as Kozai-Lidov oscillations (e.g. \citealt{cheetham:2018}).

Using Gemini-South/GPI and Keck/NIRC2 data, a giant exoplanet was recently confirmed to be orbiting the binary star HD~143811~AB (HIP 78663). The planet is at a separation of $\sim$59 au and has a mass of $\sim$5.6 M$_{\textrm{Jup}}$ (\citealt{Jones2025}, submitted ApJL). In this \textcolor{black}{companion paper} we characterize the stellar binary host using a new analysis of archival and catalog data. We combine a new orbit fit with SED fitting of the system to derive the individual masses and orbital properties of the binary, in order to understand HD~143811~AB~b in the broader context of planet formation within both single stars and binaries.

\section{HD~143811~AB}

HD~143811~AB, before being determined to be a binary, was identified as a member of the Sco-Cen star forming region \citep{dezeeuw:1999}. In particular, the star was assigned to Upper Scorpius by \citet{dezeeuw:1999}, \citet{pecaut:2012}, and \citet{galli:2018}. \citet{luhman:2020}, however, rejected HD~143811~AB as an Upper Sco member when conducting an Upper Sco census based on \textit{Gaia} DR2 data \citep{gaiadr2}, instead classifying it as part of ``the remainder of Sco-Cen." This detailed membership analysis was based on UVW velocities utilizing \textit{Gaia} DR2 proper motions and parallax and, in the case of HD~143811~AB, a radial velocity of $RV = -11.3 \pm 0.3$ km/s from \citet{chen:2011}. Given our analysis below, this radial velocity measurement likely captured the binary orbital motion, shifting the measurement of U from its true value by $\sim$10 km/s.

BANYAN $\Sigma$ \citep{gagne:2018} gives a 69.5\% chance that HD~143811~AB is a member of Upper Sco, a 30.3\% chance of being part of UCL (Upper Centaurus-Lupus), and a 0.2\% chance of being in the field. This analysis is based on $Gaia$ DR3 \citep{GaiaDR3} parallax and proper motion and the DR2 \citep{gaiadr2} radial velocity ($1.68 \pm 0.96$ km/s). Substituting the system radial velocity we derive below ($0.48 \pm 0.02$ km/s) we find similar probabilities, but with a slightly higher chance of Upper Sco membership: 72.1\%, compared to 27.8\% and 0.2\% chances of UCL and the field, respectively.

Given ages of Upper Sco of 10$\pm$3 Myr and UCL of $16 \pm 2$ Myr \citep{pecaut:2012}, and the uncertainty in the subgroup membership, we adopt an age for the system of $13 \pm 4$ Myr. This is consistent with the 2D age map of \citet{pecaut:2016}, where ages between $\sim$10--17 Myr are close to the 2D location of HD~143811~AB.

More recently, HD~143811~AB was identified as a spectroscopic binary from analysis of high resolution spectroscopy. \citet{zakhozhay:2022} obtained 15 epochs of FEROS spectra, and note the target is a double lined spectroscopic binary (SB2) with a period between 20-80 days with a companion mass in the stellar regime. We utilize these archival FEROS data, as well as archival spectra from additional epochs, below in our analysis of the system. Similarly, \citet{grandjean:2023} \textcolor{black}{obtained} three HARPS (High Accuracy Radial velocity Planet Searcher) spectra of the system within one day, noting the cross-correlation function of HD~143811~AB is double-peaked, suggesting a stellar binary.

\textcolor{black}{While HD 143811 AB has an infrared excess, and so likely contains a cold debris belt, the parameters of the debris disk are not currently well constrained. Analysis by \citet{chen:2011}, \citet{mcdonald:2012}, \citet{ballering:2013}, \citet{chen:2014}, and \citet{cotten:2016} all find evidence for a disk using either Spitzer or ALLWISE data (or both). However, the estimated disk radius varies from 8.5 AU \citep{cotten:2016}, to 23.8 AU \citep{chen:2014}, to 90 AU \citep{chen:2011}. Going forward, a reanalysis of the disk properties of this system will be necessary, especially to consider the binary nature of the host star, which was not known at the time of these past studies. Of particular interest will be whether the disk is between the orbits of the stellar binary ($\sim$0.2 AU) and the planet ($\sim$60 AU), or if instead the disk lies beyond the planet's orbit.}

%(More debris disk speculation here?)

\section{Analysis}
\subsection{Data}
HD~143811~AB was observed by the MPG/ESO 2.2-meter telescope/FEROS (Fiber-fed Extended Range Optical Spectrograph, \citealt{kaufner:1999}) on 22 epochs between 17 Apr 2018 and 18 Apr 2022.  These R$ \sim$48,000 spectra clearly show two sets of stellar lines, identifying the system as an SB2, and in many epochs the two sets of lines are clearly separated. At other epochs, near conjunction, the velocities of the two stars are more comparable, and the lines overlap.  The combination of multiple years of coverage and the $\sim$1 day cadence of many observations allows excellent time sampling of the orbital motion. Since FEROS obtains spectra of both the science target and a wavelength calibration lamp simultaneously, the spectra have exceptional wavelength stability with radial velocity precision as low as $\sim$ 6 m/s \citep{zakhozhay:2022}. We retrieve the reduced spectra from the ESO Phase 3 archive.

The two stellar components of HD~143811~AB are near equal brightness, and are close enough ($\sim$1 mas) that the data contain combined spectra of both stars in all epochs. Nevertheless, the large RV semi-amplitude of each star in the system ($\sim$25 km/s) and the high spectral resolution of FEROS result in significant separation of spectral lines at some epochs.

\subsection{Spectral Fitting}
%\subsubsection{Data}

We extract radial velocities and $v$sin($i$) for each star by generating a model of the combined spectrum and comparing to the observations, using a grid of PHOENIX stellar atmosphere models \citep{hauschildt:1999} from \citet{Husser2013}. We explore the parameter space of the model with Markov Chain Monte Carlo (MCMC) as implemented in the \textit{emcee} package \citep{emcee}. Our model consists of 11 parameters: RV offsets for both the primary and the secondary, effective temperature for both components, log($g$) for both components, $v$sin($i$) for both components, system metallicity, a single flux ratio between the two stars, and an additional broadening parameter $\sigma$. This last term accounts for resolution differences between FEROS and PHOENIX, but also broadening due to atmospheric seeing. The PHOENIX grid is linearly interpolated in effective temperature, metallicity, and log($g$), generating two spectra, one for each component of the binary. The resolution of the PHOENIX template is degraded to match that of the FEROS spectra using a Gaussian kernel with a width set to $\sigma$, and the model spectrum is further broadened using a $v$sin($i$) kernel. Next, we scale the model spectrum for the B component using the flux ratio and then combine the two spectra before normalizing to the continuum level of the data. We perform the fit on 9 regions ($\sim$3-8 \r{A} in width) of the spectrum with well-modeled narrow lines (mostly Fe I and Fe II lines) between 6,000 and 8,000 \r{A}. Error bars are assigned to each spectral channel based on the channel-to-channel variation seen in continuum regions, and then inflated by a factor of two to account for spectral covariance and model mismatch. Our final error bars are about 3\% for each spectral channel.

\textcolor{black}{We identify 11 epochs with well separated lines by visual inspection. In particular, we classify spectra based on whether the flux returns to the continuum level when moving from a line belonging to HD 143811 A to the same line in HD 143811 B. We find our reference lines to be well-separated when the two stars have a relative radial velocity $\Delta$RV~$\gtrsim$~20 km/s, and are more heavily blended otherwise. This is consistent with many of these lines having a width of $\sim$0.4 \r{A}, or about $\sim$20 km/s.  For these 11 epochs we perform a fit with uniform priors in $v$sin($i$), RV, T$_{\rm eff}$, metallicity, and $\sigma$. These broad priors have $-50$~km/s~$\le$~RV~$\le$~50~km/s, 4000~K~$\le$~T$_{\rm eff}~\le$~8000~K, 1~km/s~$\le$~$v$sin($i$)~$\le$~10~km/s, $-0.5$~$\le$~Z~$\le$~0.5, 0.1~$\le$~FR~$\le$~1.0, and 0.01~$\le\sigma\le$~100. Since our method does not reliably fit log($g$) (see Appendix), we select priors for log($g$) of each star based on the grids from \citet{Baraffe:2015} for 10-20 Myr stars at 1.3M$_\odot$ and 1.15M$_\odot$. We use Gaussian priors on log($g$) of $4.1 \pm 0.1$ and $4.2 \pm 0.1$ for the primary and secondary, respectively. As expected, the posteriors on log($g$) match the priors.}

\textcolor{black}{For epochs where the relative RVs are smaller than 20 km/s (11 epochs) and the lines mostly overlap, using wide priors results in poor fits with degeneracies between the multiple parameters. Therefore, for these blended epochs we use narrower priors on the stellar parameters, based on the fit results to epochs with more separated lines. We sum the posteriors on the stellar parameters from the 11 widely-separated epochs (to better account for systematic errors). This produces Gaussian priors of T$_{\rm eff,A} = 6439 \pm 132$K, T$_{\rm eff,A} = 5900 \pm 157$K, $v$sin($i$)$_A = 6.8 \pm 0.6$ km/s, $v$sin($i$)$_B = 2.8 \pm 1.4$ km/s, Z$=-0.18 \pm 0.1$, FR$ = 0.76 \pm 0.05$, and $\sigma = 33 \pm 3.6$, which we use when fitting the blended epochs. The radial velocity prior remains the same: $-50$~km/s~$\le$~RV~$\le$~50~km/s.}  

\begin{figure}
    \centering
    \includegraphics[width=0.75\linewidth]{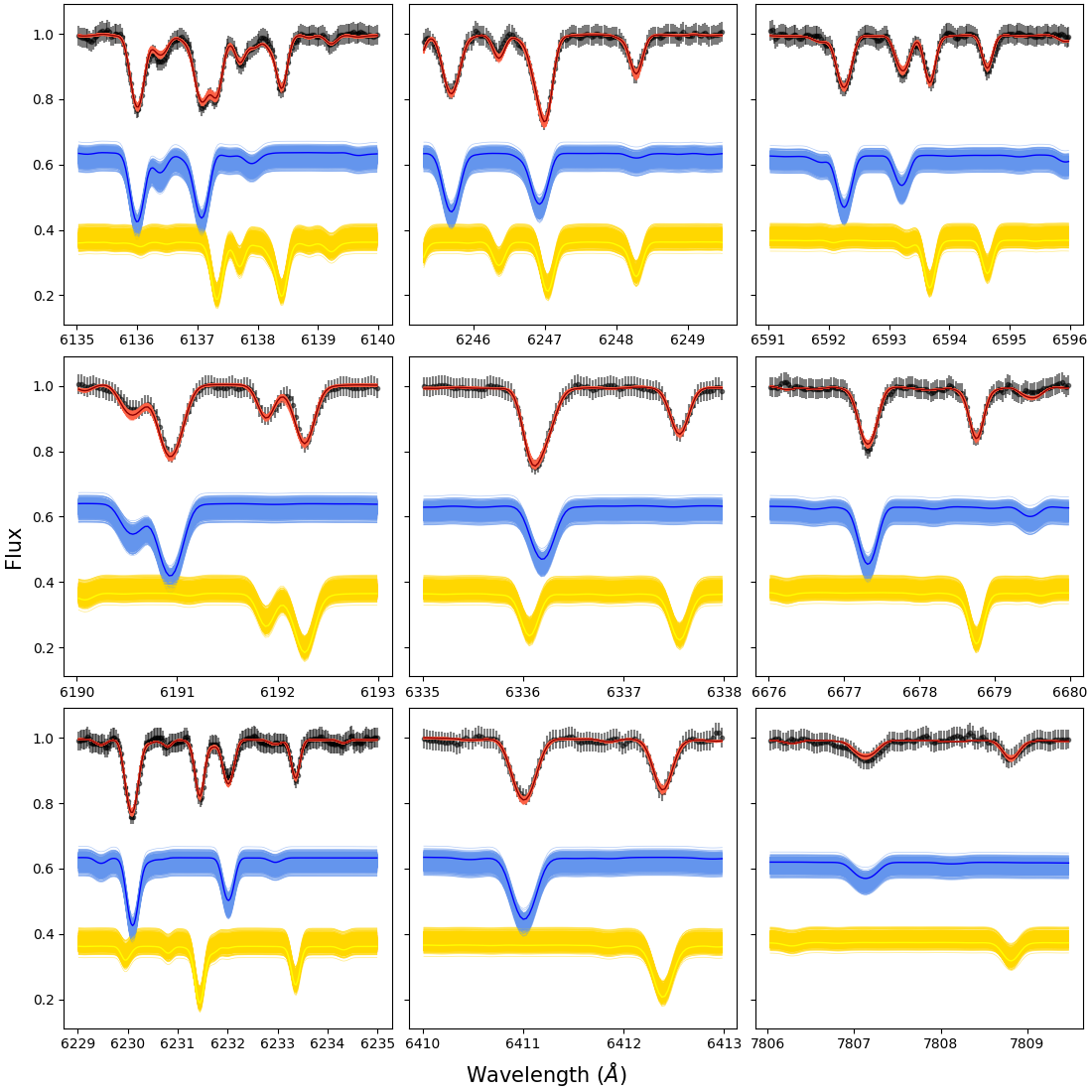}

    \caption{\textcolor{black}{FEROS spectra of HD~143811~AB  taken on 2018-04-22} (black points with error bars), showing two sets of well-resolved lines. Red tracks are draws from the posterior (lowest chi-square solution in dark red) of a two-star model that best fits the data. The blue and yellow lines are draws from single star models for A and B, respectively. Given the high resolution and SNR of these data we are able to recover both the radial velocities of both components as well as $v$sin($i$).}
    \label{fig:spectralfit}
\end{figure}

\textcolor{black}{The temperatures and metallicities of HD 143811 A and B are not well constrained by our spectral fitting method (see Appendix). To ensure that our RVs are accurate, we perform a second round of spectral fitting with more precise priors on temperature and metallicity from the SED fit (described below in Section~\ref{sec:stellar_param}), which in turn depends on the mass ratio from the orbit fit (Section~\ref{sec:orbit_fit}). The RVs from this second round of spectral fitting are used for a second round of orbit fits. There is a negligible change in the RVs and the mass ratio between the initial and final round of fitting. For example, the RVs for the first epoch changed from $12.47 \pm 0.12$ km/s and $-13.37 \pm 0.13$ km/s to $12.46 \pm 0.12$ km/s and $-13.35 \pm 0.13$, while the posterior on mass ratio changed from $0.8852 \pm 0.0025$ to $0.8856 \pm 0.0026$. As a result, we conclude our results are self-consistent after this second round of fitting, and adopt the second set of RVs and orbital parameters going forward. The final RVs for the 22 epochs are given in Table~\ref{tab:radveltable}. 
Figure~\ref{fig:spectralfit} shows all 9 segments of the spectrum from one epoch, with FEROS data in black, and fits in red. Overall there is good agreement between model and data, though as expected there are mismatches between the model and data for some lines.
}

\textcolor{black}{While the parameters derived from each epoch generally agree, final values for T$_{\rm eff}$, $v$sin($i$), and metallicity are calculated from the summed 1D posteriors across all epochs to better incorporate systematic errors. Error bars are taken to be the standard deviations of the combined posterior, with an additional error of 107.2 K added in quadrature to the uncertainty on temperature. The derivation of this additional error term, from a comparison of our method to literature results on 8 stars, is presented in the Appendix. Stellar parameters from this fit, and additional fits described below, are given in Table~\ref{tab:startable}.}

%Details of the priors are provided in Table \ref{tab:startable}. An example of the posterior distributions for each parameter for one epoch is presented in the Appendix in Figure \ref{fig:bigcorner}.

 \subsection{Orbit Fitting}\label{sec:orbit_fit}

 We perform an orbit fit (using the same Metropolis-Hastings MCMC method as \citealt{nielsen:2020}) to the SB2 based on our measured velocities, fitting for the RV semi-amplitude of both stars, period, eccentricity, argument of periastron, epoch of periastron passage, and system RV. Uniform priors are assumed on all parameters, with eccentricity bounded between $0 \le e \le 0.95$. Ten MCMC chains are run in parallel for $10^7$ steps each, with every 100th step saved. We assess convergence both by visual inspection of the posteriors and by considering the Gelman-Rubin statistics, which is below 1.001 for all parameters. Since the data contain both high-cadence observations (15 epochs over two weeks) and long baselines (four years between the first and last epoch) the constraints on the orbital parameters are excellent.

\begin{figure}
    \centering
    \includegraphics[width=1.0\linewidth,trim=0 3cm 0 8cm]{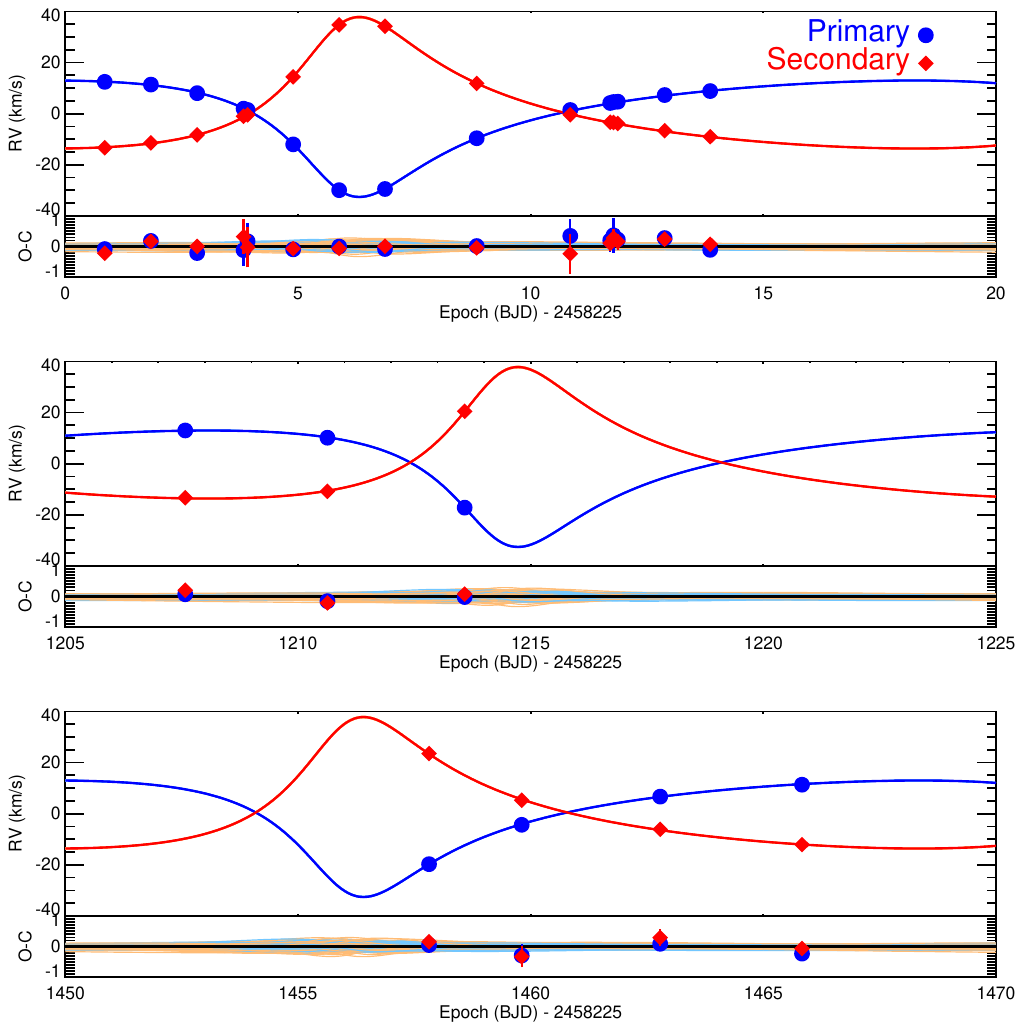}
    
    \caption{\textcolor{black}{Radial velocity measurements} of both HD~143811~A and HD~143811 B with orbits drawn from the posteriors, with the darkest blue and red lines representing the lowest $\chi^2$ orbit. The Observed-Calculated (O-C) panels show the residuals with respect to the lowest $\chi^2$ orbit (black line at 0), as well as the offset of other draws from the posterior with respect to the best-fitting orbit. The $\sim$18 day orbit is an excellent fit to the data, and the combination of multiple observation over 2 weeks and a multi-year baseline strongly constrain the orbital parameters.}
    \label{fig:rvfit}
\end{figure}

\begin{figure}
    \centering
    \includegraphics[width=1.0\linewidth,trim=0 3cm 0 8cm]{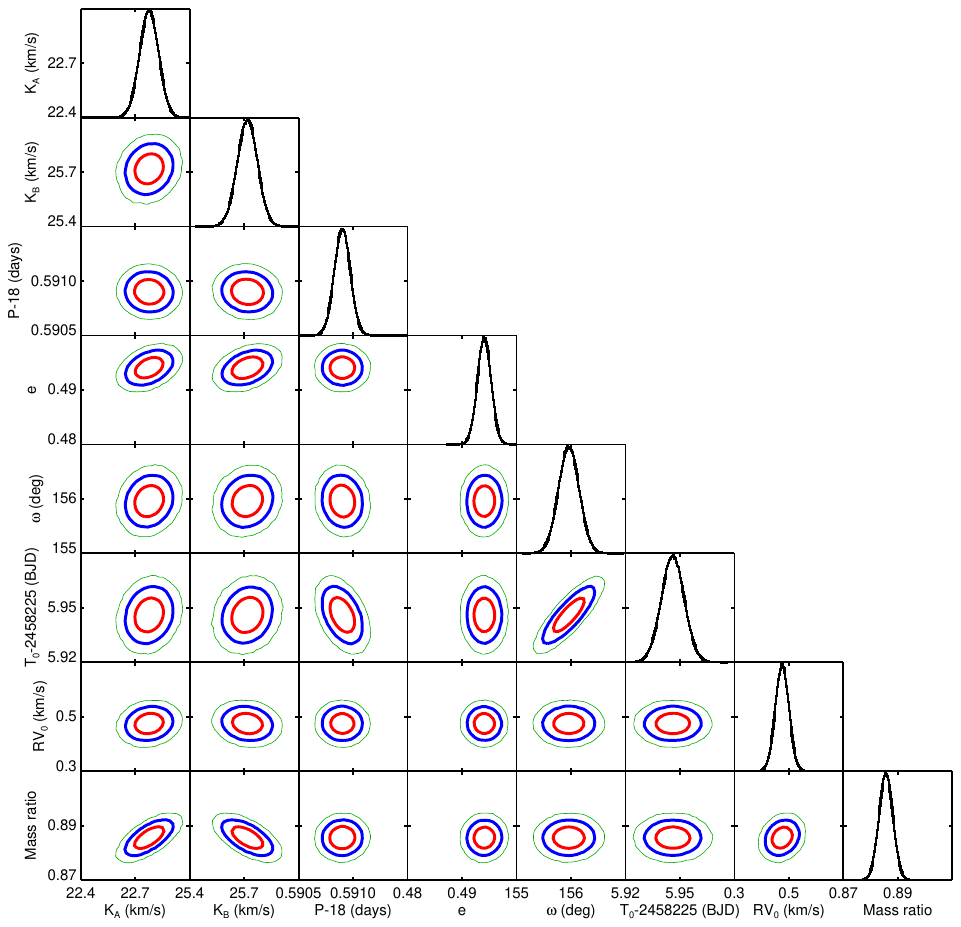}
    
    \caption{\textcolor{black}{Posterior on the} SB2 radial velocity orbit of HD~143811~AB. The orbit is very well constrained by the FEROS data, with orbital period known to \textcolor{black}{6} second precision. Direct measurement of both semi-amplitudes allows a measurement of mass ratio at the sub-1\% level: $0.886 \pm 0.003$.}
    \label{fig:rvcorner}
\end{figure}

Figure~\ref{fig:rvfit} shows the RVs of both components and draws from the posterior. The best-fitting orbit has a reduced $\chi^2$ close to unity, indicating our RV error bars are appropriate. The posteriors on this fit are shown in Figure~\ref{fig:rvcorner}. Period is constrained to approximately six seconds: $P = 18.59090 \pm 0.00007$ days, and mass ratio constrained to less than a percent: $q = 0.886 \pm 0.003$. The significant eccentricity of the orbit is well-determined, $e = 0.4941 \pm 0.0013$, and the $\sim$90\% mass ratio is consistent with the similarity of the two spectra, as well as the similar depths of the lines.

\subsection{Stellar Parameters}\label{sec:stellar_param}

  The SB2 orbit does not provide direct measurements of the individual masses, semi-major axis, or inclination angle. Nevertheless, with the unresolved photometry of the system and an estimate for the binary mass ratio, we can derive model-dependent values of the stellar parameters. We follow the same procedure outlined in \citet{nielsen:2019} for stars with $B-V > 0.35$, using a joint evolutionary-atmospheric model to fit the blended photometry of the system to estimate the masses of the two components. As in \citet{nielsen:2019}, we use a modified version of the MESA Isochrones and Stellar Tracks (MIST) model grid \citep{Dotter:2016,Choi:2016} combined with ATLAS9 model atmospheres \citep{Castelli:2003}. With this approach, we can estimate the spectral energy distribution (SED) of the two stars given the masses of the two components ($M_1$, $M_2$) and an age ($t$), metallicity ($[M/H]$), parallax ($\varpi$), extinction ($A_V$) for the system. The joint evolutionary-atmospheric model allows us to translate these parameters into the observable SED of the system, the two components being unresolved and blended together due to their very small angular separation. This can then be used to fit the observed optical photometry from {\it Gaia} \citep{GaiaDR3} and near-infrared photometry from 2MASS \citep{Skrutskie:2006}. While this system has a debris disk, the observed excess is at $\sim$24 microns, and so the disk should not impact the bluer photometry. The observed and modeled SED is shown in Figure~\ref{fig:sedfit}, while the triangle plot for this fit is given in the Appendix in Figure~\ref{fig:sedfit_triangle}.

 %With the stellar effective temperatures and metallicity calculated from the SED fit in hand, we rerun out spectral fits for each epoch. This time, we adopt the posteriors for effective temperature and metallicity from the SED fit as priors for the SED fit. All other priors remain the same. 

\begin{deluxetable}{ccccc}[h]
    \caption{\textcolor{black}{Radial Velocities of HD~143811~A and B. Widely separated epochs are where the two sets of lines are distinct from each other, which corresponds to $\Delta RV \gtrsim 20 km/s$}\label{tab:radveltable}}
    \tablecolumns{3}
    \tablehead{Epoch (BJD) & UTC & RV$_A$ (km/s) & RV$_B$ (km/s) & Widely Separated?}
    \startdata
2458225.8481768593 & 2018-04-17 08:21:22.480645 & 12.457$\pm$0.125 & -13.349$\pm$0.129 & Y\\
2458226.840259357 & 2018-04-18 08:09:58.408435 & 11.382$\pm$0.125 & -11.445$\pm$0.136 & Y\\
2458227.8335984894 & 2018-04-19 08:00:22.909485 & 8.041$\pm$0.135 & -8.285$\pm$0.152 & N\\
2458228.831037526 & 2018-04-20 07:56:41.642248 & 1.962$\pm$0.508 & -0.994$\pm$0.578 & N\\
2458228.9179943185 & 2018-04-20 10:01:54.709121 & 1.462$\pm$0.592 & -0.456$\pm$0.65 & N\\
2458229.8972283546 & 2018-04-21 09:32:00.529836 & -12.01$\pm$0.13 & 14.416$\pm$0.133 & Y\\
2458230.8849802855 & 2018-04-22 09:14:22.296671 & -29.906$\pm$0.123 & 34.725$\pm$0.126 & Y\\
2458231.870725756 & 2018-04-23 08:53:50.705322 & -29.459$\pm$0.125 & 34.198$\pm$0.129 & Y\\
2458233.8412399422 & 2018-04-25 08:11:23.131009 & -9.609$\pm$0.119 & 11.851$\pm$0.132 & Y\\
2458235.851854297 & 2018-04-27 08:26:40.211249 & 1.438$\pm$0.558 & -0.442$\pm$0.644 & N\\
2458236.710698977 & 2018-04-28 05:03:24.391623 & 4.191$\pm$0.381 & -3.368$\pm$0.264 & N\\
2458236.7799807256 & 2018-04-28 06:43:10.334693 & 4.561$\pm$0.561 & -3.431$\pm$0.306 & N\\
2458236.8734692587 & 2018-04-28 08:57:47.743948 & 4.693$\pm$0.349 & -3.851$\pm$0.262 & N\\
2458237.8811232424 & 2018-04-29 09:08:49.048146 & 7.296$\pm$0.196 & -6.654$\pm$0.21 & N\\
2458238.8585430426 & 2018-04-30 08:36:18.118877 & 8.852$\pm$0.128 & -9.01$\pm$0.136 & N\\
2459432.578191188 & 2021-08-06 01:52:35.718657 & 12.993$\pm$0.122 & -13.352$\pm$0.132 & Y\\
2459435.6336258487 & 2021-08-09 03:12:25.273326 & 10.156$\pm$0.132 & -10.816$\pm$0.131 & Y\\
2459438.5826384313 & 2021-08-12 01:58:59.960468 & -17.199$\pm$0.113 & 20.502$\pm$0.147 & Y\\
2459682.817435831 & 2022-04-13 07:37:06.455784 & -19.745$\pm$0.117 & 23.526$\pm$0.137 & Y\\
2459684.8084221077 & 2022-04-15 07:24:07.670107 & -4.349$\pm$0.382 & 5.28$\pm$0.344 & N\\
2459687.784224547 & 2022-04-18 06:49:17.000859 & 6.681$\pm$0.249 & -6.123$\pm$0.274 & N\\
2459690.8318373566 & 2022-04-21 07:57:50.747606 & 11.342$\pm$0.126 & -12.101$\pm$0.145 & Y\\
    \enddata
    
\end{deluxetable}

With our combined model in hand, we use an MCMC-based approach for parameter estimation. We use Gaussian priors for the age ($t=13\pm4$\,Myr; synthesizing the age for the US and UCL sub-groups from \citealp{pecaut:2016}), parallax ($\varpi=7.3065\pm0.0204$\,mas; \citealp{GaiaDR3}), metallicity ($[M/H]=-0.05\pm0.11$\,dex; \citealp{Nielsen:2013}), and mass ratio ($q=0.885\pm0.003$). From this, we find individual masses of the two stars of $M_A = 1.30^{+0.03}_{-0.05}$ M$_\odot$ and $M_B = 1.15^{+0.03}_{-0.04}$ M$_\odot$, although the two masses are highly covariant given the tight constraint on the mass ratio.

Based on the sum of these individual masses and the orbital period, we find a semi-major axis (for the orbit of B around A) of $0.1854^{+0.0014}_{-0.0024}$ au. Combined with the eccentricity of the orbit and the distance to the system, apastron should be $\sim$2 mas. While this is too close for traditional imaging, it should be reachable with an interferometer like VLTI/GRAVITY \citep{gravity:2017}. This semi-major axis is a factor of $\sim$300 smaller than the planet separation; such a large difference would be expected for a stable orbit of a planet around a binary star.

\textcolor{black}{An RV orbit does not constrain the orbital inclination angle, but we can use the SED-derived masses to estimate the inclination angle of the stellar binary orbit. In particular, semi-amplitude is a function of inclination angle, as well as masses of both components, period, and eccentricity. By combining posteriors from the orbit fit for semi-amplitude for HD~143811~A, period, and eccentricity with posteriors on the two masses from the SED fit we then obtain a posterior on orbital inclination angle. We repeat this process with the semi-amplitude of HD~143811~B to obtain a second posterior on the orbital inclination angle. As expected for self-consistent results, both semi-amplitudes return similar values of inclination angle: $i_A = 22.9^{+0.3}_{-0.2} $$ \degree$ and $i_B = 22.8^{+0.3}_{-0.2}$$ \degree$. We adopt the first value as the model-dependent inclination angle of the system. Since radial velocity measurements cannot distinguish between clockwise and counterclockwise orbits, a value of $i = 157.1^{+0.2}_{-0.3} $$ \degree$ is also possible.}

\begin{figure}
    \centering
    \includegraphics[width=\linewidth]{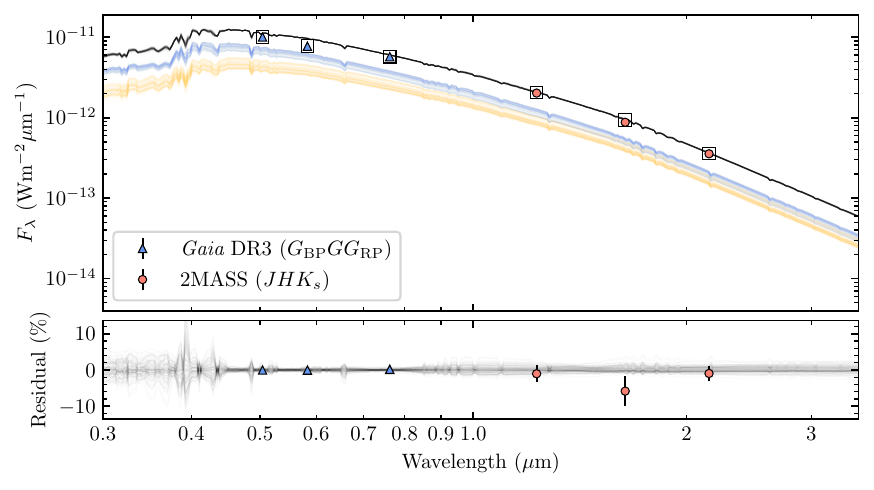}
    \caption{The results of a two-star fit to the unresolved optical and near-infrared photometry of the system from {\it Gaia} and 2MASS. Plotted are the spectral energy distribution of the blended system (black) and the two components (blue, yellow) drawn from the posterior distributions estimated using MCMC. Also shown are the synthetic photometry of the blended system derived from the median SED (black squares), and the observed photometry (circles).  \label{fig:sedfit}}
    
\end{figure}
%We use the FEROS spectra to derive physical and orbital properties of the HD~143811~AB binary system and the individual stars. 

 \begin{deluxetable}{cccccccc}[h]
    \caption{\textcolor{black}{Properties of HD~143811~AB. Adopted parameters are indicated with a Y. We use an age for the system of $13 \pm 4$ Myr based on the uncertain Upper Sco or UCL membership, though the posterior from the SED fit favors somewhat older ages. Temperature and metallicity are adopted from the SED fit rather than the spectral fit.} \label{tab:startable}}
    \tablecolumns{4}
    \tablehead{Parameter & HD~143811 A & HD~143811 B & Reference & & & Adopted?}
    \startdata
    Age (Myr) & \multicolumn{2}{c}{$13 \pm 4$ }  & 1 &  & &Y& \\
    Distance (pc) & \multicolumn{2}{c}{136.9 $\pm$ 0.4} & 2 & & &Y&\\
     & & \\
         \multicolumn{3}{l}{\textit{SED Fitting}}  && \textit{Prior} & \textit{Prior Reference} &\\
    $[M/H]$ (dex) & \multicolumn{2}{c}{$-0.15^{+0.07}_{-0.08}$}  & 1& $0.05\pm0.11$ & 3 & Y\\
    Age (Myr) & \multicolumn{2}{c}{$20.9^{+1.8}_{-1.7}$}  & 1& $13 \pm 4$ & 1 & N\\
%    $\bar{\omega}$ (mas) & \multicolumn{2}{c}{$0.16^{+0.10}_{-0.09}$}  & 1\\
    A$_V$ (mag) & \multicolumn{2}{c}{$0.16\pm0.01$}  & 1& Uniform & 1&Y\\
    Mass (M$_\odot$) & $1.30^{+0.03}_{-0.05}$ & $1.15^{+0.03}_{-0.04}$ & 1 & $q=0.885\pm0.003$ & 1 (orbit fit) & Y\\
    T$_{\rm eff}$ (K)& $6751^{+207}_{-148}$ & $6349^{+122}_{-133}$ & 1&& & Y\\
%    log g & $4.1\pm0.1$ & $4.2\pm0.1$ & 1\\
    %v$sin($i$) (km/s) & $6.8\pm{0.6}$ & $2.8\pm{1.4}$ & 1\\
         & & \\
    \multicolumn{3}{l}{\textit{Spectral Fitting}} && \textit{Prior} & \textit{Prior Reference} & \\
   $[M/H]$ (dex) & \multicolumn{2}{c}{$-0.078\pm{0.037}$}& 1 & $-0.15^{+0.07}_{-0.08}$ & 1 (SED Fit) & N\\
    $T_{\rm eff}$ (K)& $6572\pm{139}$ & $6134\pm{127}$ & 1&$6751^{+207}_{-148}$, $6349^{+122}_{-133}$ & 1 (SED Fit) & N\\
    %log g & $4.1\pm0.1$ & $4.2\pm0.1$ & 1 &$4.1\pm0.1$& 3\\
    $v$sin($i$) (km/s) & $6.73^{+0.70}_{-0.82}$ & $2.89^{+1.41}_{-1.29}$ & 1& Uniform, 1--10 & 1 & Y\\
     Flux Ratio & \multicolumn{2}{c}{$0.75^{+0.14}_{-0.05}$}  & 1 &  Uniform, 0.01--1 & 1 & Y\\
         & & \\
    \multicolumn{3}{l}{\textit{Orbit Fitting}}  && \textit{Prior} & \textit{Prior Reference}\\
    $K$ (km/s) & 22.78 $\pm$ 0.05 & 25.72 $\pm$ 0.06 & 1& Uniform & 1 & Y \\
    Period (days) &  \multicolumn{2}{c}{$18.59090 \pm 0.00007$} & 1& Uniform & 1 & Y\\     
    Eccentricity &  \multicolumn{2}{c}{$0.4941 \pm 0.0013$} & 1& Uniform, 0--0.95 & 1 & Y\\
    $\omega$ (deg) & $ 155.96  \pm 0.19$ & $335.96 \pm 0.19$ & 1& Uniform & 1 & Y\\
    T$_0$ (BJD) &  \multicolumn{2}{c}{$2458230.946 \pm 0.006$} & 1& Uniform & 1 & Y\\
    RV$_0$ (km/s) &  \multicolumn{2}{c}{$0.48 \pm 0.02$} & 1& Uniform & 1& Y\\
    $ q= \frac{M_B}{M_A}$ &  \multicolumn{2}{c}{$0.886 \pm 0.003$} & 1&&& Y\\
         & & \\
    \multicolumn{3}{l}{\textit{Combined Results}}  && & \\
%    Mass (M$_\odot$) & $1.30^{+0.03}_{-0.05}$ & $1.15^{+0.03}_{-0.04}$ & 1\\
    Semi-major axis (AU) & \multicolumn{2}{c}{$0.1854^{+0.0014}_{-0.0024}$} & 1 &&& Y \\
    Orbit Inclination Angle (deg) & \multicolumn{2}{c}{$22.9^{+0.3}_{-0.2}$ or $157.1^{+0.2}_{-0.3}$} & 1&&& Y \\
    \enddata
    \tablerefs{1: this work, 2: \cite{GaiaDR3}, 3: \cite{Nielsen:2013}.}% 4: \cite{Baraffe:2015}}
    
\end{deluxetable}

 %For epochs where the relative RVs are smaller than 20 km/s (11 epochs) and the lines mostly overlap, using wide priors results in poor fits with degeneracies between the multiple parameters. While these epochs are not suitable for additional measurements of the stellar parameters, they can still be used to extract RVs of both stars. For the epochs with more overlapping lines we use narrower priors, based on the fit results to epochs with more separated lines. While the radial velocity prior remains $-50$~km/s~$\le$~RV~$\le$~50~km/s, we adjust the priors for the other parameters for the overlapping epochs. We adopt Gaussian priors from our adopted posteriors on the stellar parameters, based on the fitting to epochs with more widely-separated lines. The measured RVs for each epoch are reported in Table~\ref{tab:radveltable}.

As expected for young F-stars we detect significant lithium absorption in both stars near 6707.8 \r{A}. \textcolor{black}{We follow the fitting procedure described in \cite{Peck2025}, but we adapt it to measure lithium equivalent widths for both components of the stellar binary rather than a single star by scaling the continuum based on the flux ratio. We measure equivalent widths of $74.3\pm2.7$m\r{A} and $105.0\pm3.5$m\r{A} for the primary and secondary respectively. Since we do not have accurate B-V values for either component, we do not attempt to calculate age posteriors using \texttt{BAFFLES} \citep{BAFFLES}. However, the measured equivalent widths are consistent with those observed for F stars in $\beta$ Pic, IC 2602, $\alpha$ Per, and the Pleiades.} We do not observe signification emission cores in Calcium H and K or in H$\alpha$.

\section{Discussion}

\subsection{Physical Parameters}

The relatively large eccentricity of HD~143811~AB ($\sim$0.5) is not unexpected given the orbital period of $\sim$18 days. \citet{torres:2021}, examining Pleiades binaries, find that tidal circularization is most prominent for orbital periods below $\sim$7 days. The binary period itself is typical of FGK binaries, though the peak of the distribution is closer to $10^5$ days ($\sim$300 years), the log-normal distribution is quite broad (\citealt{Duquennoy:1991}, \citealt{Raghavan:2010}). The discovery of a planet around such a binary is partially the result of an important selection effect: wider-separation binaries ($\sim$0.1-1") would be excluded from the GPIES target list, or aborted once the binary became visible. This is due both to the expected lack of dynamical space for a planet with similar orbital separation to a stellar binary, and the practical effect that a star not occulted by the coronagraph would saturate the detector. As a result, binary stars identified as planet hosts by direct imaging surveys tend to either have very small orbits (so that both stars fit comfortably under the coronagraph), or very large orbits (so that the other star is well outside the field of view of the detector).

There is a significant offset between the temperatures we derive for the two components from fitting the SB2 spectrum \textcolor{black}{($6572 \pm 139$ and $6134 \pm 127$ K)} compared to fitting the SED and the mass ratio ($6751^{+207}_{-148}$ and $6349^{+122}_{-133}$K). The SED results are systematically \textcolor{black}{$\sim$200 K} higher than the spectral fitting results; the  \textcolor{black}{$\sim$1 $\sigma$ confidence intervals touch for HD~143811~A and B.} In the SED fit, there is a strong covariance between the extinction and both temperatures, with lower values of extinction corresponding to lower temperatures. There is a range of extinction values across Sco-Cen, both across subgroups (the mean $A_V$ is 0.73 and 0.17 for Upper Sco and UCL, respectively, \citealt{degeus:1989,wright:2018}) and within subgroups. \citet{pecaut:2016}, for example, find multiple members of both Upper Sco and UCL with large extinctions ($A_V \sim 1$), as well as stars with extinctions as low as $A_V = 0$. Setting the extinction in the SED fit to $A_V = [0.00, 0.05, 0.10]$ moves the median temperatures of A and B to $[6530,6564,6636]$~K and $[6104,6216,6297]$~K, respectively; the lowest of these values are more in line with the spectral fitting results. 

Alternatively, there may be a systematic offset in the spectral fitting when fitting an SB2 compared to fitting single stars. A significant issue is the incompleteness of models, especially lines that appear in the model but not the data or vice versa. These are visible when comparing FEROS data to the PHOENIX models, and we chose the spectral regions for the fit to isolate well-modeled lines. However, this can become more important when the RV offset of an SB2 results in a well-modeled line in one star overlaps with a poorly-modeled line in the other.

\textcolor{black}{When applying a single star version of the spectral fit to well characterized stars, we find that for temperatures above $\sim$ 6000 K, our method systematically underpredicts temperature and does not reliably measure metallicity, though it does recover accurate radial velocities (see Appendix). Therefore, we adopt the temperatures and metallicities from the SED fit rather than the spectral fit.} While the masses of the two components have some covariance with temperature in the SED fit, this effect is relatively small. As a result, our derived masses of the two components ($1.30^{+0.03}_{-0.05}$ and $1.15^{+0.03}_{-0.04}$ M$_\odot$) are generally better constrained than the temperatures.

\subsection{Sco-Cen Membership}

The rotation and metallicity of these stars are generally consistent with Sco-Cen, as expected. We find relatively low $v$sin($i$) for both A and B: $6.8 \pm 0.6$ km/s and $2.9 \pm 1.2$ km/s, respectively. Compared to other Sco-Cen stars of similar $B-V$ color from \citet{grandjean:2023} (their Figure 1b), these values are at the low end of their HARPS Sco-Cen sample, which span $v$sin($i$) from $\sim$3 to $\sim$100 km/s. It is possible the two components of HD~143811~AB are being viewed more pole-on compared to other Sco-Cen members, or that tidal forces affected their rotational evolution. Slower overall rotation would be consistent with the lack of strong calcium emission we observe in the spectra. Our derived metallicity for the system from the \textcolor{black}{SED fit, $-0.15^{+0.07}_{-0.08}$}, is within the distribution of $0.0 \pm 0.18$ measured by \citet{grandjean:2023}.

The system radial velocity of $RV_0 = 0.48 \pm 0.02$ km/s, obtained from our orbit fit, is $\sim 1 \sigma$ lower than the $Gaia$ DR2 velocity of $1.68 \pm 0.96$ km/s, and significantly higher than the \citet{chen:2011} value of $-11.3 \pm 0.3$ km/s. This large difference is to be expected given the semi-amplitude of each component of $\sim$25 km/s. By combining our system radial velocity with $Gaia$ DR3 astrometry we find space motions of the HD~143811~AB system of $(U,V,W) = (-2.68 \pm 0.04, -18.10 \pm 0.05, -4.68 \pm 0.02)$. This places HD~143811~AB $(2.1,1.1,0.9) \sigma$ from the central UVW of Upper Sco, as given by \citet{wright:2018}. The same calculation places the system slightly further from the central UVW of UCL, also from \citet{wright:2018}: $(1.6,2.6,0.9) \sigma$. This is in line with the BANYAN $\Sigma$ results, when we combine $Gaia$ DR3 astrometry with our radial velocity we find 72.1\% probability of Upper Sco membership, 27.8\% for UCL, and 0.2\% for the field. The SED fit seems generally more consistent with UCL than Upper Sco, with the $20.9^{+1.8}_{-1.7}$ Myr age from that fit closer to the $\sim$16 Myr UCL than the $\sim$10 Myr Upper Sco. Similarly, a lower value of extinction (which would bring the SED and spectral fit temperatures more in line with each other) would also suggest UCL membership. Going forward, better age constraints on HD~143811~AB will allow for a more definitive determination of subgroup membership.

It is also interesting to compare HD~143811~AB~b to another Sco-Cen directly imaged planet: HD 95086 b \citep{rameau:2013}. Both planets formed in the same association, are about the same age, orbit at roughly the same distance ($\sim$60 au), and are about the same mass ($\sim$5 M$_{\textrm{Jup}}$). The major difference is the host star: HD 95086 is a $\sim$1.6 M$_\odot$ A8 single star, while HD~143811~AB is a close binary containing a 1.30 M$_\odot$ mid-F star and a 1.15 M$_\odot$ late-F star. A key question is whether both planets formed at their present location, or if they moved outward after formation, either from early interactions with the protoplanetary disk or due to gravitational scattering. Gravitational scattering is expected to play an important role in another Sco-Cen planet orbiting a tight binary, HD 106906 b \citep{bailey:2014}, which is at a projected separation of $\sim$740 au. This very large separation is thought to result from gravitational interactions between the planet, the inner binary, and an additional Sco-Cen star that passed close to this system \citep{derosa:2019,nguyen:2021}. If both HD~143811~AB~b and HD 95086 b are close to their formation location it might suggest that wide-separation giant planet formation around close binary stars proceeds in a similar way as around single stars. The role of gravitational scattering can be tested with future orbit monitoring of the planet. In particular, if HD~143811~AB~b has a more eccentric orbit than most directly imaged planets \citep{bowler:2020} a scattering event might be more likely compared to in-situ formation. Further analysis of the binary orbit of HD~143811~AB, in particular the extent to which it is aligned (or not aligned) with the planet's orbit, will also provide dynamical constraints on the origins of HD~143811~AB~b.

With the current data it is not yet possible to robustly determine if the three-dimensional orbits of both the inner stellar binary and the outer planet are coplanar. In particular, the radial velocity data on the stellar binary provides no information on the position angle of nodes of the inner orbit. Similarly, with only a few degrees of orbital motion, the planet's position angle of nodes is not well-determined, where despite four peaks in the posterior there is non-zero probability for all values between 0 and 360$\degree$ (\citealt{Jones2025}, submitted ApJL). Without inclination angle and position angle of nodes measurements for both orbits, mutual inclination cannot be directly computed. Nevertheless, coplanarity cannot be ruled out with the current observations, with a model-dependent inclination angle of the binary orbit of \textcolor{black}{$i_{AB} = 22.9^{+0.3}_{-0.2} $$ \degree$}, and an inclination angle for the planet's orbit of $i_{b} = 38 \pm 17 \degree$ (\citealt{Jones2025}, submitted ApJL). Thus, both inclination angle posteriors overlap at the $\sim 1 \sigma$ level. Retrograde orbits are still possible, however, if \textcolor{black}{$i_{AB} \approx 157$} is the actual inclination of the stellar binary. Additional orbital monitoring of the planet will allow a more precise inclination angle measurement and a quantitative comparison of the two inclination angles (e.g. \citealt{kennedy:2013}). Measurements of both orbits' position angles of nodes, which for the inner binary would likely require VLTI/GRAVITY or \textit{Gaia} DR4, would allow for a more direct test of coplanarity.

\section{Conclusions}

We report the characterization of the HD~143811~AB binary, the host of the directly imaged planet HD~143811~AB~b. We extract RVs of both components from archival spectra at 22 epochs and derive orbital parameters, consistent with an eccentric $\sim$18 day orbit, and a mass ratio of $\sim$0.9. Combining with the age of the system, unresolved photometry, and evolutionary models we derive masses of $M_A = 1.30^{+0.03}_{-0.05}$ M$_\odot$ and $M_B = 1.15^{+0.03}_{-0.04}$ M$_\odot$ for the two components.

Future observations of the HD~143811~AB system will refine the orbital parameters of both orbits. While the RV orbit of HD~143811~AB is very well constrained, interferometry of the system will make it possible to resolve the $\sim$1 mas visual orbit of the binary, and so directly measure the full 3D orbit. In addition, $Gaia$ DR 4 will contain individual astrometric scans, which may have the precision to detect the $\sim$0.1 mas photocenter orbit of the binary. Similarly, additional orbital monitoring of the planet will better constrain its 3D orbit, allowing a direct measurement of the mutual inclination of the two orbits. In addition to aiding 3-body modeling of the system, these improved orbits will place valuable constraints on the formation mechanism and dynamical evolution of this system.

An improved exoplanet orbit will be especially interesting when coupled with future long wavelength observations of the star, such as from JWST MRS. More accurately determining the debris disk's radius and extent, especially given the physical parameters of the stellar binary, will better inform the dynamics of the disk, and its interactions with the planet.

HD~143811~AB~b joins a growing list of directly imaged planets in binary systems including 51 Eri b \citep{macintosh:2015} and  HD 106906 b \citep{bailey:2014, derosa:2019}. With additional detections, it will become feasible to determine how the occurrence rate of wide-separation giant planets depends on binary properties, and how it compares to the occurrence rate for single stars. This comparison has traditionally been difficult, given the relatively low occurrence rate of wide-separation giant planets and the tendency of imaging surveys to screen for binaries (e.g. \citealt{bonavita:2016}). Improvements to high contrast imaging, such as the upcoming Gemini Planet Imager 2.0 \citep{chilcote:2024}, will enable better sensitivity to giant planets orbiting single stars, orbiting one star in wide binaries, and orbiting both stars in close binaries. $Gaia$ DR4 will also allow a comparison of closer-in giant planet populations between binaries and single stars, especially in cases where astrometry from both components of the binary can be extracted.

\begin{acknowledgments}

\textcolor{black}{We would like to thank the referee for their insightful and helpful comments, which significantly improved this paper.}

Based on observations obtained at the international Gemini Observatory, a program of NSF NOIRLab, which is managed by the Association of Universities for Research in Astronomy (AURA) under a cooperative agreement with the U.S. National Science Foundation on behalf of the Gemini Observatory partnership: the U.S. National Science Foundation (United States), National Research Council (Canada), Agencia Nacional de Investigación y Desarrollo (Chile), Ministerio de Ciencia, Tecnología e Innovación (Argentina), Ministério da Ciência, Tecnologia, Inovações e Comunicações (Brazil), and Korea Astronomy and Space Science Institute (Republic of Korea).

This publication makes use of data products from the Two Micron All Sky Survey, which is a joint project of the University of Massachusetts and the Infrared Processing and Analysis Center/California Institute of Technology, funded by the National Aeronautics and Space Administration and the National Science Foundation

Based on observations collected at the European Organisation for Astronomical Research in the Southern Hemisphere under ESO programme(s) 0101.A-9012(A), 0107.A-9004(A), 0109.A-9014(A). This research has made use of the SIMBAD database, CDS, Strasbourg Astronomical Observatory, France. This research has made use of the VizieR catalogue access tool, CDS, Strasbourg Astronomical Observatory, France (DOI : 10.26093/cds/vizier).

This work has made use of data from the European Space Agency (ESA) mission $Gaia$
(https://www.cosmos.esa.int/gaia), processed by the Gaia Data Processing and Analysis Consortium (DPAC, https://www.cosmos.esa.int/web/gaia/dpac/consortium). Funding for the DPAC has been
provided by national institutions, in particular the institutions participating in the $Gaia$ Multilateral Agreement.

Based on observations obtained with the Apache Point Observatory 3.5-meter telescope, which is owned and operated by the Astrophysical Research Consortium.

A.E.P., W.R., E.L.N., and A.A. are supported by NASA grant 80NSSC21K0958 and NSF grant AST 2510959. E.L.N. and A.J.R.W.S. are supported by NASA grant 21-ADAP21-0130. B.L.L. acknowledges support from the National Science Foundation Astronomy \& Astrophysics Postdoctoral Fellowship under Award No. 2401654. A.S. is supported by the National Science Foundation Graduate Research Fellowship under Grant No. 2139433. Any opinions, findings, and conclusions or recommendations expressed in this material are those of the author(s) and do not necessarily reflect the views of the National Science Foundation. 

\end{acknowledgments}

\facilities{$Gaia$, $Hipparcos$, $Max Planck:2.2m$, ARC}

\software{\texttt{astropy} \citep{astropy:2013, astropy:2018, astropy:2022}, \texttt{SciPy} \citep{2020SciPy-NMeth}, \texttt{NumPy} \citep{numpy}, \texttt{emcee }\citep{emcee}}

\appendix 
\label{append}
\textcolor{black}{We investigate the accuracy of our stellar parameter fitting using archival FEROS spectra of 8 well-studied F and G stars. We compare our stellar parameters to those measured \citet{worley2012} using these same FEROS spectra. Using the spectral fitting method described in this work, modified to model a single star instead of an SB2, we measure the effective temperatures and metallicities of these stars. We do not fit for log($g$), because we find that log($g$) values extracted from our method are significantly different from the literature values. This could be the result of the PHOENIX model grids being too coarsely sampled in log($g$), our choice of lines not constraining log($g$), or limitations in our science spectra such as the spectral resolution or continuum correction, all of which could prevent our method from constraining log($g$). Since gravity is covariant with metallicity and temperature, we place a prior on log($g$), rather than attempt to measure it from our spectral fits. We use a log($g$) prior for each star of a Gaussian, from the value and error reported by \citet{worley2012}.}

\begin{figure}[h]
    \centering
    \includegraphics[width=0.45\linewidth]{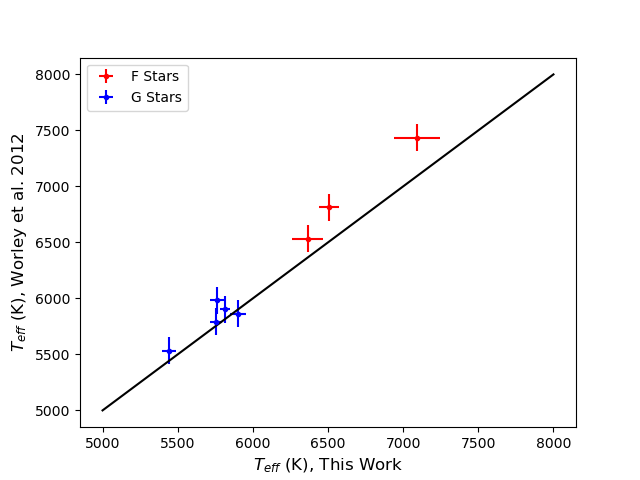}
    \includegraphics[width=0.45\linewidth]{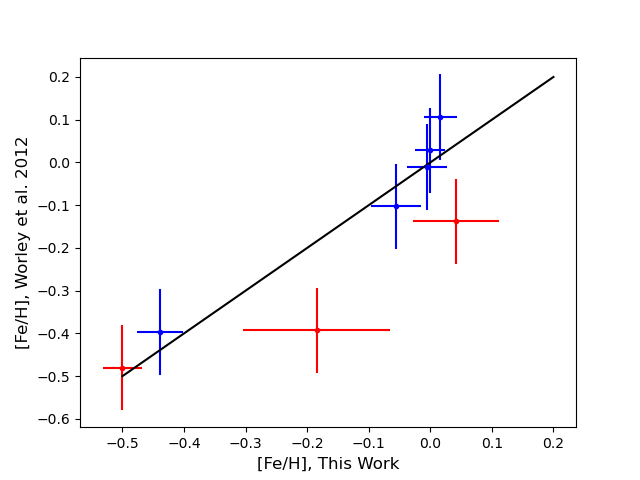}
    
    \caption{\textcolor{black}{We are generally able to accurately recover effective temperature and metallicity when applying our method to well-characterized F and G stars from \citet{worley2012} However, the results for stars hotter than $\gtrsim$6000 K show a systematic offset in temperature and a larger scatter in metallicity compared to G stars. The solid black line is the 1-to-1 line, representing identical results.}}
    \label{fig:teff_z_comparison}
\end{figure}

\textcolor{black}{Our measured effective temperatures and metallicities for these 8 stars are shown in Figure \ref{fig:teff_z_comparison}. The effective temperature residuals show a systematic offset at temperatures greater than $\sim 6000$ K, with our method recovering higher effective temperatures compared to \citet{worley2012}. The metallicity residuals do not show a systematic offset, however our metallicity measurements for the F stars are significantly more discrepant with the \citet{worley2012} values compared to the better-matching G stars. The residuals in effective temperature have a median of $-129.5$ K and a standard deviation of $123.3$ K, while the metallicity residuals have a median of $-0.007$ dex and a standard deviation of $0.100$ dex. The offset in effective temperature is likely due to the wavelength ranges we used in our analysis, but further analysis is required to confirm this. Therefore, we adopt the effective temperatures and metallicities for HD 143811 A and B (both with effective temperatures above 6000 K) from the SED analysis rather than our spectral modeling. }

\textcolor{black}{For effective temperature the spread in the residuals is not consistent with the expected errors, as calculated by adding our measurement errors and those from \citet{worley2012} in quadrature. While we do not adopt these values, in order to report them with more accurate errors we use the residuals to measure an error correction. We assume that our errors are underestimated, but do not account for the systematic offset in this step, and model the additional noise as Gaussian, using maximum likelihood estimation to determine this additional error. We find that we need to add $107.2$ K quadrature to our effective temperature uncertainties, which we utilize when reporting temperatures for HD 143811 A and B from the spectral fits. Regardless of temperature and metallicity measurements, the measured RVs of both these 8 comparison stars and HD 143811 A and B are consistent across fits.}

\begin{figure}
    \centering
    \includegraphics[width=1.0\linewidth]{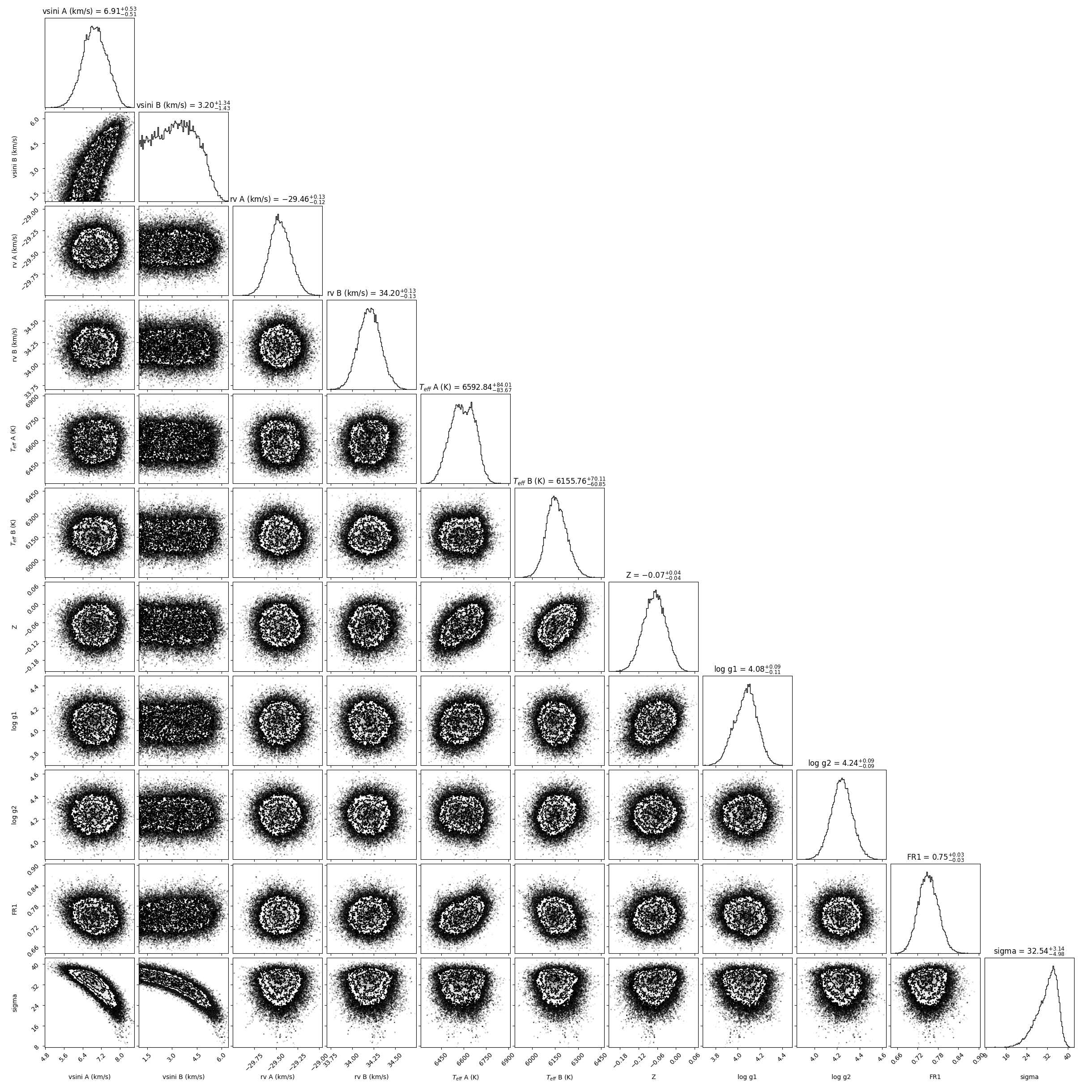}
    \caption{The posterior distribution for the binary fit to the 2018-04-18 FEROS spectrum, fitting for stellar parameters and radial velocities of both components of the binary.}
    \label{fig:bigcorner}
\end{figure}

\begin{figure}
    \centering
    \includegraphics[width=0.8\linewidth]{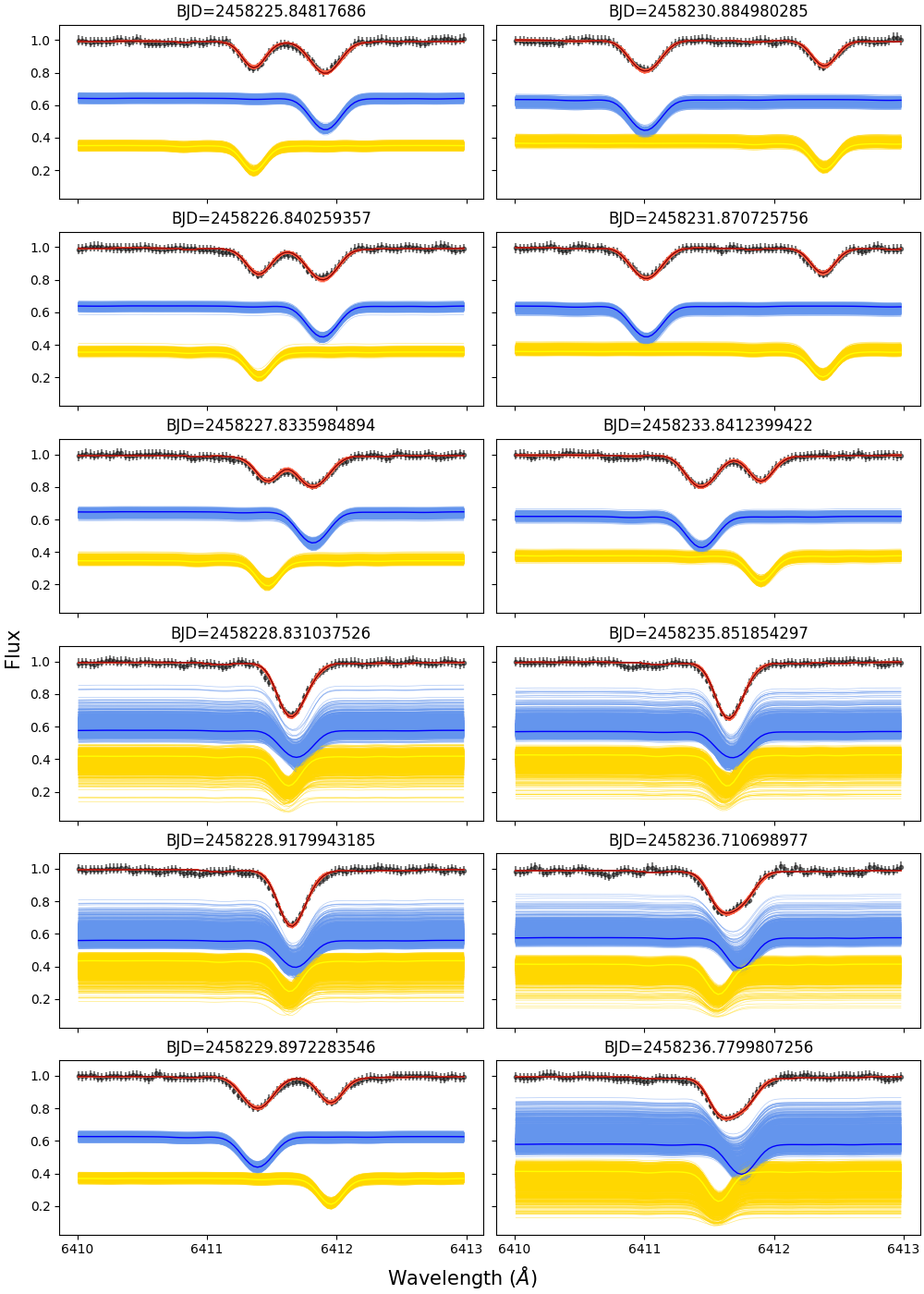}
    \caption{\textcolor{black}{A single wavelength region used for the spectral fitting for all 22 epochs. Black points with error bars are the data. Red tracks are random draws from the posterior for the two star model fit. Blue and yellow tracks are random draws from the posterior for single star models for the primary and secondary, respectively. Our method finds reliable RVs for both epochs with widely-separated lines and epochs with blended lines.}}
    \label{fig:allepochs1}
\end{figure}

\begin{figure}
    \centering
    \includegraphics[width=0.8\linewidth]{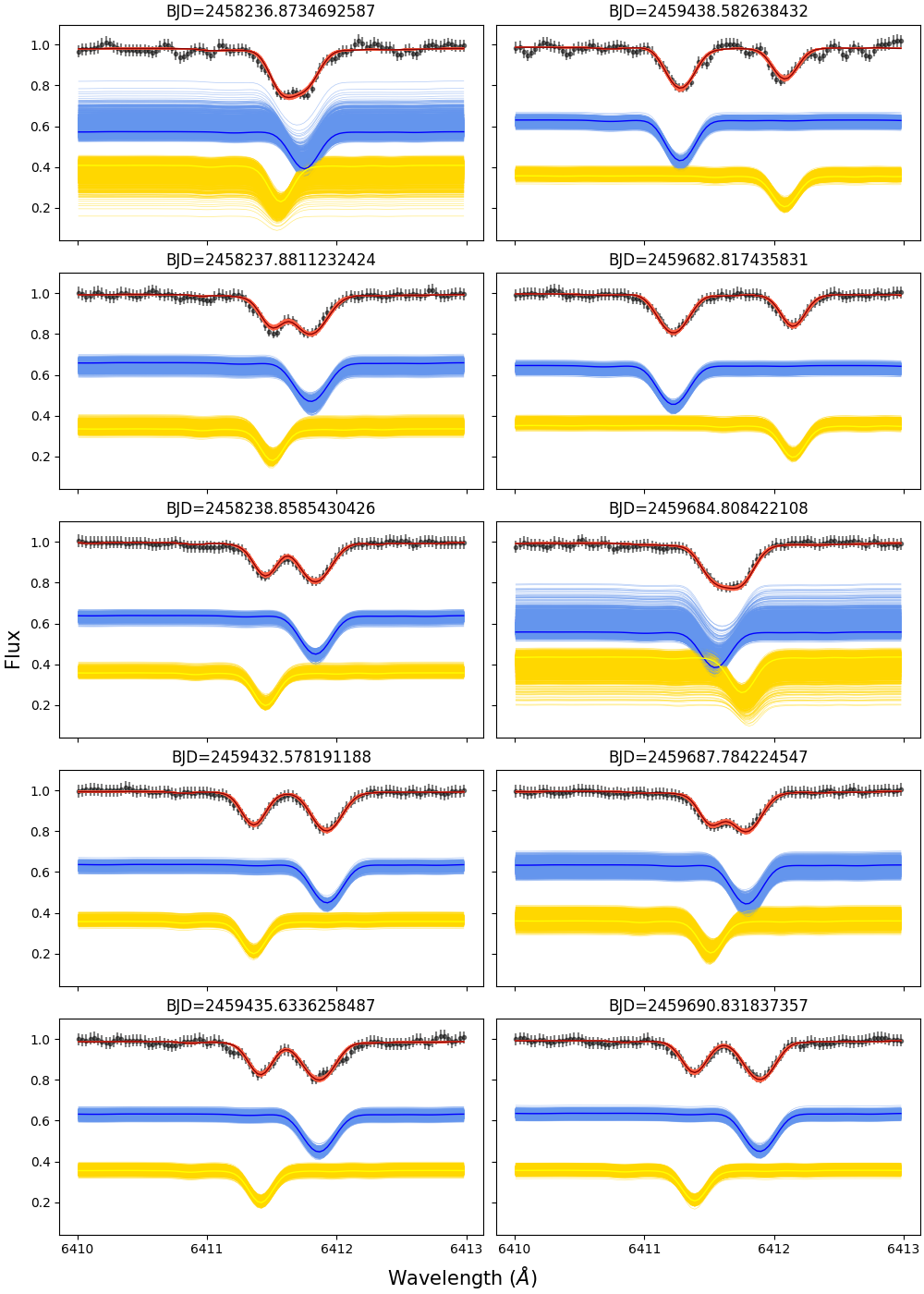}
    \caption{Continued from Figure \ref{fig:allepochs1}.}
    \label{fig:allepochs2}
\end{figure}

\begin{figure}
    \centering
\includegraphics[width=\linewidth]{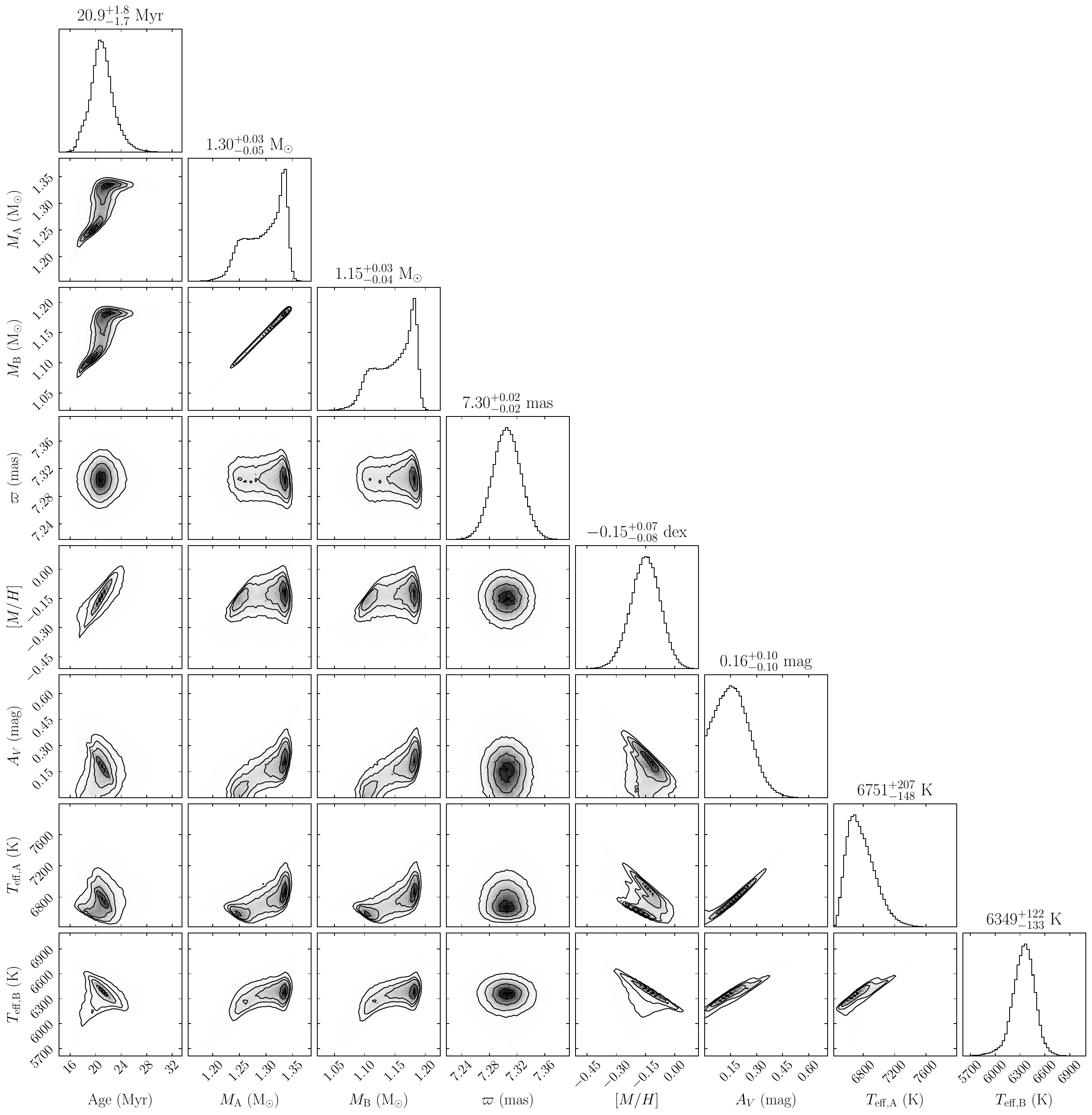}
    \caption{The posterior distribution for the two-star fit to the unresolved photometric measurements, giving the six fitted parameters and their covariances. The effective temperature, a derived parameter, of both components is also shown. \label{fig:sedfit_triangle}}
    
\end{figure}

%\section{.}
%.

\bibliography{ms}{}
\bibliographystyle{aasjournalv7}

\end{document}